\documentclass[conference]{IEEEtran}
\IEEEoverridecommandlockouts
\usepackage{cite}
\usepackage{amsmath,amssymb,amsfonts}
\usepackage{algorithmic}
\usepackage{graphicx}
\usepackage{textcomp}
\usepackage{xcolor}
\usepackage{enumerate}
\usepackage{caption}
\usepackage{enumitem}
\def\BibTeX{{\rm B\kern-.05em{\sc i\kern-.025em b}\kern-.08em
    T\kern-.1667em\lower.7ex\hbox{E}\kern-.125emX}}
\begin{document}

\title{Secure and Privacy-preserving Network Slicing in 3GPP 5G System Architecture}

\author{\IEEEauthorblockN{Xiangman Li, Miao He, and Jianbing Ni} \IEEEauthorblockA{Department of Electrical \& Computer Engineering, Queen's University, Kingston, Canada K7L 3N6\\}
 Email: \{xiangman.li, miao.he, jianbing.ni\}@queensu.ca}

\maketitle

\begin{abstract}
Network slicing in 3GPP 5G system architecture has introduced significant improvements in the flexibility and efficiency of mobile communication. However, this new functionality poses challenges in maintaining the privacy of mobile users, especially in multi-hop environments. In this paper, we propose a secure and privacy-preserving network slicing protocol (SPNS) that combines 5G network slicing and onion routing to address these challenges and provide secure and efficient communication. Our approach enables mobile users to select network slices while incorporating measures to prevent curious RAN nodes or external attackers from accessing full slice information. Additionally, we ensure that the 5G core network can authenticate all RANs, while avoiding reliance on a single RAN for service provision. Besides, SPNS implements end-to-end encryption for data transmission within the network slices, providing an extra layer of privacy and security. Finally, we conducted extensive experiments to evaluate the time cost of establishing network slice links under varying conditions. SPNS provides a promising solution for enhancing the privacy and security of communication in 5G networks.
\end{abstract}

\section{Introduction}


Network slicing, a vital technology in the era of 5G and beyond, is a revolutionary approach that enables the dynamic creation of multiple virtual networks on top of a single physical network infrastructure. Network slicing comprises three primary components: the Radio Access Network (RAN) slice, the core network slice, and the management and orchestration framework \cite{netslicing2}. RAN refers to the portion of a mobile network that connects user devices to the core network, the core network is responsible for managing data traffic, security, and network services, while the management and orchestration framework oversees the coordination, automation, and optimization of virtualized network resources and services. network slicing revolutionizes the way resources are allocated in a network, combining partitioning, isolation, customization, and dynamic resource allocation to enhance performance and meet diverse requirements for use cases and mobile users. 

Despite the appealing advantages, network slicing brings new challenges, one of which is privacy leakage of mobile users \cite{khan2019survey}. First, since multiple entities can share the same physical network infrastructure, there is a risk of potential data leakage or unauthorized access. Without adequate security measures, sensitive information transmitted through one network slice becomes vulnerable to potential access or interception by other slices that share the same underlying infrastructure \cite{lu20195g}. Second, when multiple slices coexist on the sam physical infrastructure, the traffic patterns, resource allocation, or even unique characteristics of certain services within a particular slice can be observed or inferred by other slices. Third, each network slice is identified with an ID, which is linked to the service type that a mobile user can access. The exposure of network slide ID can leak the specific service or application being used within the network slide for a mobile user. The leakage of service types can have various implications. For example, it can reveal competition information or sensitive user preferences. Even worse, a malicious adversary may gain insights into the vulnerabilities or weakness of a service or application, facilitating targeted attacks or unauthorized access.

To prevent privacy leakage in network slicing for 3GPP 5G systems, Ni et al. \cite{Efficientandsecureservice} proposed the first privacy-preserving network slicing protocol that enables users to construct secure data channels between users and remote servers, while achieving anonymous authentication with servers with the authorization of both 5G core networks and servers. Since then, secure and privacy-preserving network slicing has received increasing attentions. For example, Porambage et al. \cite{porambage2019secure} proposed a secure keying scheme for network slicing architecture that can protect data against tampering attacks and key-compromise impersonate attacks and privacy protection of mobile users. Hum et al. \cite{mun2021secure} proposed a secure V2V communication based on 5G by utilizing network slicing to guarantee different features of V2X services and security requirements in network slicing. Zhang et al. \cite{zhang2021flexible} designed an anonymous network slicing method for cloud radio access network that achieves authentication of emerging 5G service and privacy preservation for mobile users. Sathi et al. \cite{sathi2020novel} proposed privacy-preserving authentication protocols that prevent the exposure of users' service access behavior to device-to-device communication peers and third-party service providers. However, these methods are hard to be integrated into 3GPP 5G system architecture due to complex cryptographic operations and huge delay on data delivery.

In this paper, we propose a secure and privacy-preserving network slicing protocol (SPNS) that resolves security and privacy issues brought by network slicing. Specifically, SPNS enables users to select network slices, with measures in place to prevent curious RAN nodes or external attackers from obtaining full slice information, including service types and network slice ID. Also, it implements end-to-end encryption for data transmission within the network slices based on the key agreement for building secure channels and support the 5G core network to authenticate the connected RANs in dual connectivity, which is an important feature of 5G networks, offering several benefits such as improved coverage, increased reliability, and faster data rates. With privacy preservation and dual connectivity, the mobile users' sensitive information can be well protected, including preferences, location information, and accessing services.

\textsf{Outline.} The remainder of this paper is organized as follows: Section \textsc{II} comprehensively reviews the 5G network slicing architecture, delving into its background and fundamental concepts. In Section \textsc{III}, we formalize the security threats, and design goals of our proposed solution. The detailed construction of our novel design is presented in Section \textsc{IV}, followed by an in-depth security analysis in Section \textsc{V}. Finally, Section \textsc{VI} showcases the performance evaluation of our design, and Section \textsc{VII} offers concluding remarks.


\section{Background}
\subsection{ End-to-End Architecture of 5G Network Slicing}

The end-to-end architecture of 5G network slicing enables network operators to create and manage network slices that are customized to meet the specific needs of different applications and users. It consists of three main components: the radio access network (RAN), the core network, and the service layer. Each of these components plays a critical role in enabling network operators to create and manage network slices that are customized to meet the specific needs of different applications and users \cite{Survey}.

The RAN is responsible for providing wireless connectivity to end-users. In 5G network slicing, the RAN is designed to support multiple network slices, each with its own set of requirements and characteristics.

The core network is responsible for managing the data traffic between the RAN and the service layer. In 5G network slicing, the core network is designed to support multiple network slices, each with its own set of requirements and characteristics. To enable this, the core network is divided into three main components: the control plane, the user plane, and the network slice selection function (NSSF) \cite{Efficientandsecureservice}. In the 5G network architecture, the Access and Mobility Function (AMF), Session Management Function (SMF), and Unified Data Management (UDM) all belong to the Control Plane components. They are responsible for handling control signaling related to network connectivity, mobility, session management, and user data \cite{3GPP.23.501}. The Control Plane deals with signaling information related to the establishment, maintenance, and teardown of network connections, as well as management tasks related to mobility, security, and QoS. The User Plane, on the other hand, primarily carries the actual user data (e.g., voice, video, text) and is not involved in control signaling. In a 5G network, the User Plane Function (UPF) is a typical User Plane component. At the same time, the NSSF is a key component of the core network in 5G network slicing. Its main function is to select the appropriate network slice for each application and user \cite{Overview}. The NSSF takes into account various factors, such as application requirements, user preferences, and network conditions, to determine the best network slice to use.

The service layer provides the applications and services that run on the network \cite{3GPP.23.501}. In 5G network slicing, the service layer is designed to support multiple network slices, each with its own set of requirements on delay, throughput, and QoS.

\subsection{Network Slicing Process}

\begin{figure}
    \centering
    \includegraphics[width=0.5\textwidth]{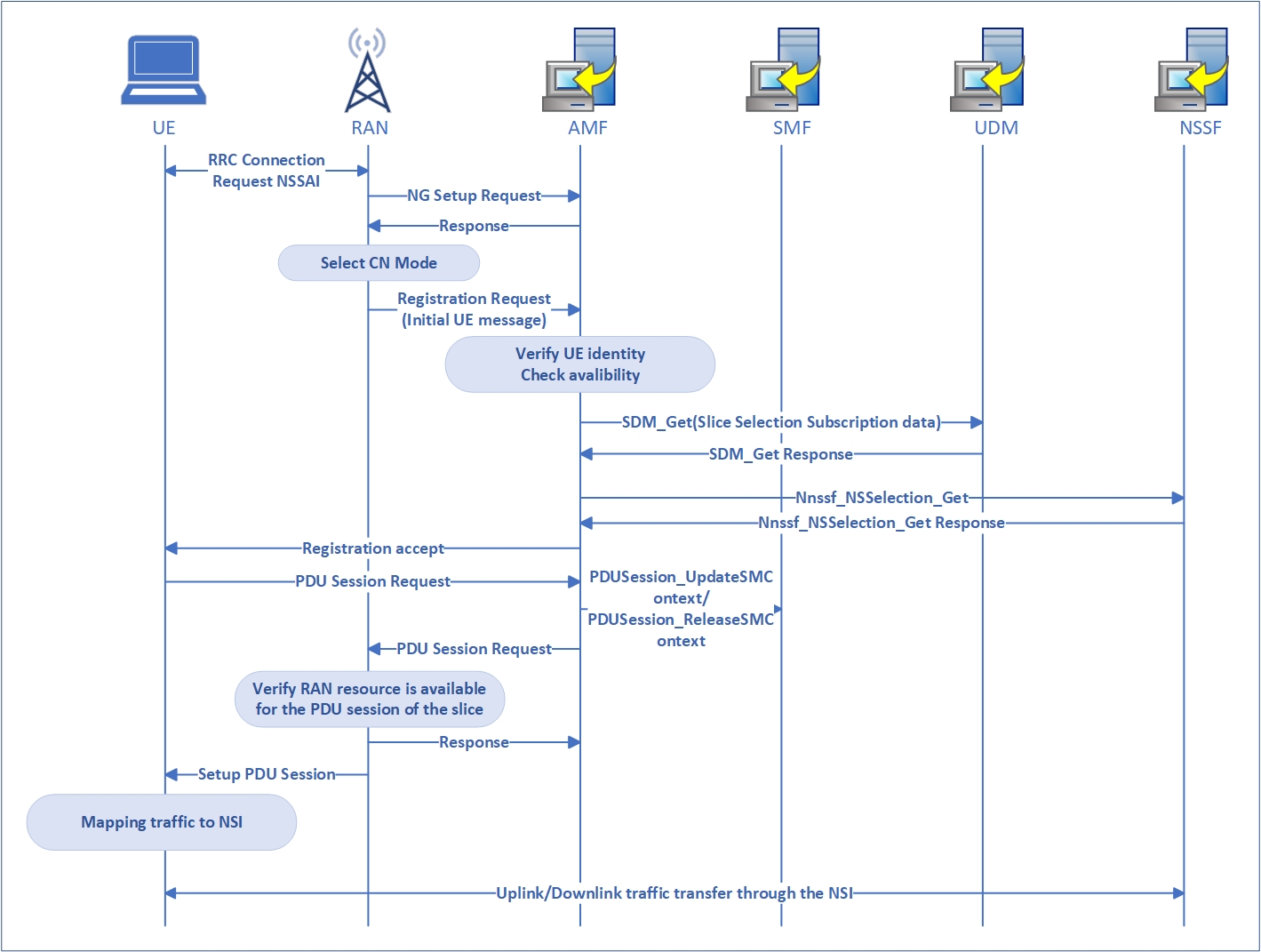}
    \caption{Network Slicing Processing}
    \label{networkSlicing}
\end{figure}

Base on 3GPP TS 23.502\cite{3GPP.23.502}, the main process of 5g network slicing is shown as Fig. \ref{networkSlicing} \cite{Overview}.

The establishment of a Radio Resource Control (RRC) connection between a User Equipment (UE) and a Public Land Mobile Network (PLMN) entails the UE indicating the desired Network Slice Selection Assistance Information (NSSAI) for the establishment of the desired network slice. The UE selects the NSSAI value based on its specific requirements on delay, throughput, and other QoS, following which the Radio Access Network (RAN) identifies a suitable Access and Mobility Management Function (AMF) and forwards the request to the AMF. The AMF then communicates with the User Data Management (UDM) function, which manages the subscribed network information. The AMF sends an SDM\_Get message to the UDM, and the UDM sends an SDM\_Get response to the AMF, finalizing the UDM selection.

Subsequently, the AMF determines the approved Single Network Slice Selection Assistance Information (S-NSSAI) based on the subscribed information. In case the requested NSSAI is not approved, the AMF seeks the assistance of the Network Slice Selection Function (NSSF), which selects an appropriate network slice instance and informs the AMF of the permissible NSSAI. Upon initiation of a Protocol Data Unit (PDU) session, the customized user plane functions of the selected network slice handle the processing of user data traffic by the control plane network functions.

In addition, to ensure optimal service, these network slices are continuously monitored and managed. This allows network operators to offer tailored services to different user groups, thereby supporting a wide range of applications and use cases.

\section{Syntax}
In this section, we will introduce the protocol overview, security threats, and our design goal.
\subsection{Overview}
We consider dual connectivity in 5G system architecture, which allows for seamless switching between two RANs for mobile users, ensuring a stable and reliable connection even in areas with limited network coverage. In addition to its coverage benefits, dual connectivity also offers increased location privacy. By connecting to two different RANs, the device's location cannot be precisely determined by any single RAN, as the location information is distributed across both networks. This makes it more difficult for third parties to pinpoint the device's exact location, providing users with increased privacy and security. This is particularly important for industries that handle sensitive data, such as healthcare or finance.
We show a 5G network slicing construction design, consisting of four phase: System Initialization, Network Slicing Establish, 5G Core Connection, and Data Transmission.

    \textit{1) System Initialization. }The User Equipment (UE) selects a Master RAN and a Secondary RAN, with the Master RAN communicating with the 5G core and sending all relevant information, including the Secondary RAN information. The Secondary RAN, in turn, establishes a connection with the UE and sends the relevant service information to the Master RAN. The UE then establishes an RRC connection with the selected RAN nodes for the establishment and management of the wireless communication link. To set up the system parameters Params, the UE generates a half Diffie-Hellman key ($g^{x_{1}}, g^{x_{2}},\cdots ,g^{x_{n}}$) for each selected RAN. Each RAN maintains both a long-term key and an onion key. The long-term keys are used to sign TLS certificates, router descriptors, and directories by directory servers. The onion key, on the other hand, is used to decrypt user-issued requests to establish circuits and negotiate temporary keys.

    \textit{2) Network Slicing Establish.} $UE$ constructs Network Slicing Onion routing link, negotiating a Diffie-Hellman key with each RAN. Secondary RAN extends the link to the Master RAN, and encapsulates the network slicing information and service type received from the other RANs along with its own.

    \textit{3) 5G Core Connection.} The Master RAN in this link builds a connection with the 5G core. It encapsulates the information of all RANs and constructs an NG setup request to the 5G core. Upon receiving the request, the 5G core initiates the configuration process for network slicing established.

    \textit{4) Data Transmission.} The $UE$ can start transmitting data within the designated slice after the 5G network slicing establishment and registration process has been successfully completed, each RAN and 5G core have already negotiated network slicing service, and PDU sessions have been formed.

\subsection{ Security Threats}
The implementation of network slicing in 5G networks has introduced several security threats that can compromise user privacy, data integrity, and service availability. One of the most significant threats is eavesdropping, where adversaries can intercept data transmissions between users and the 5G core network to obtain sensitive user information and service data.
Additionally, the geographical location of users can be compromised due to the geographical location of the RAN that is serving them. Since RANs are responsible for transmitting data between the user equipment and the 5G core network, they can potentially infer the location of users based on their device's signal strength and other information.
Furthermore, the integrity of data during transmission is easily compromised, as adversaries can intercept and manipulate data sent between users and the 5G core network. The centralized nature of the 5G core network also poses a potential single point of failure, making it a high-value target for adversaries seeking to disrupt network services.
To mitigate these security threats, it is essential to provide strong security and privacy protections for users in 5G network slices.

\subsection{Design Goal}
We aim to build a protocol adopting the idea of onion routing as the approach for securing $UE$ information transmission. After the setup and network slicing instance establishment phase, each RAN and core network have already established their connections. With the use of onion routing, when $UE$ sends data, each party only knows its own information during the NSMF instance creation process. As a result, we achieve the unawareness of the core network's identity against the first RAN and the anonymity of the $UE$ identity against the second RAN. The design goals of our design are as follows:

\begin{itemize}
  \item \textit{Secure Network Slice Information:} We enable users to select network slices, with measures in place to prevent curious RAN nodes or external attackers from obtaining full slice information. Each RAN node only possesses its own complete service information and the service information disclosed by the preceding RAN node, ensuring that the network slice information remains secure and private.
  \item \textit{Anonymous Authentication:} We enable the 5G core network to authenticate all RANs, while simultaneously ensuring that the 5G core network cannot rely on a single RAN to provide services. This is achieved through the implementation of secure mechanisms that prevent the 5G core from establishing a direct connection with any individual RAN, thereby maintaining data confidentiality and preserving user privacy.
  \item \textit{End-to-End Encryption:} We implement end-to-end encryption for data transmission within the network slices, preventing unauthorized access to sensitive data and mitigating the risk of eavesdropping and MITM attacks.
\end{itemize}

\begin{figure*}[ht]
    \centering
    \includegraphics[width=\textwidth]{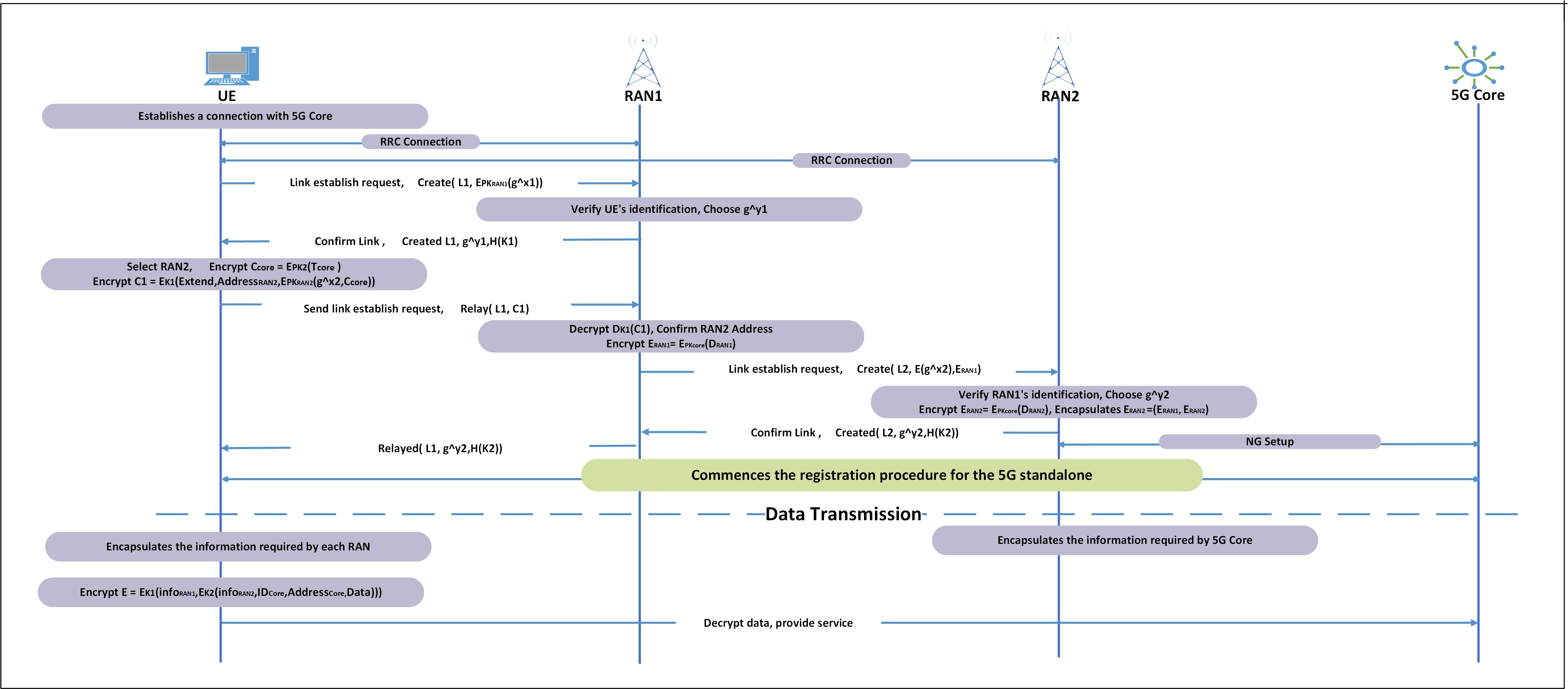}
    \caption{System Model}
    \label{System model}
\end{figure*}

\section{Construction of SPNS}
In this section, we present the overview and the detailed
construction of the proposed SPNS.

\subsection{NSI ID Design}
The Network Slice Instance Identifier (NSI ID) is a unique identifier used in 5G network slicing to distinguish different instances of network slices. The NSI ID is a 16-byte (128-bit) identifier that uniquely identifies a specific network slice instance. The Slice ID represents the identifier of a network slice, which is assigned by the Network Slice Selection Function (NSSF). It can be used by various functional modules within the network to manage and control the instance.

To achieve the Tor design in network slicing, Slicing ID and NSSAI need to separate into multiple parts. In the 5G framework, the Network Slice Instance Identifier (NSI ID) can be divided into multiple parts to represent different slices. For example, NSI ID can be divided into RAN slice, core slice, and other parts based on the application scenario or network function. This division method helps to define and manage different types of slices in the network and assign unique identifiers for each slice.

In the 5G network, the NSI ID can be represented in the Uniform Resource Name (URN) format, which is a type of identifier format that can be used to uniquely identify different types of resources, such as network slices. The URN format for the NSI ID is:
\begin{equation}
\begin{split}
urn:<slice-part><slice-part>
\end{split}
\end{equation}
Each $<slice-part>$ represents a part of the NSI ID and can be any string that represents a specific slice. The multi-part representation of the NSI ID improves the flexibility and scalability of the network to meet the needs of different users and applications.

\subsection{Detail of SPNS}
\textit{1) System Initialization. } The User Equipment ($UE$) initiates an RRC connection request by preparing a message that encompasses its identity information and the rationale for establishing the connection. This message is transmitted to Secondary RAN, denoted as $RAN1$; and Master RAN, denoted as $RAN2$. Upon receiving the RRC connection request, the RAN nodes perform a series of security and connectivity assessments to validate the $UE$'s request. If the RAN nodes ascertain the validity of the $UE$'s request, they transmit an RRC connection setup message to the $UE$. This message includes the RRC Connection Identifier designated for $UE$, as well as the radio resources and channel configurations specifically allocated to $UE$. After receiving the RRC connection setup message, $UE$ acknowledges the message and establishes a connection utilizing the provided resources. Consequently, $UE$ transits to the connected mode. $UE$ also needs to choose the AMF who is the Tor Directory Service.

In SPNS, $UE$ is designated as an Onion Proxy (OP), while each RAN function is an Onion Router (OR). As part of their roles as Onion Routers, $RAN1$ and $RAN2$ generate a router descriptor $D_{i}$ containing essential information such as the node name, gNB ID, location area, supported NSSAI with the slice ID in the NSI ID $ID_{i}$, long-term identity key, and onion key $PK_{i}$. The long-term keys serve a critical purpose in signing TLS certificates, router descriptors, and directories maintained by directory servers. Meanwhile, the onion key is employed to decrypt user-initiated requests for circuit establishment and temporary key negotiation. The onion key is typically established using robust cryptographic algorithms, such as RSA or ElGamal. In addition, 5G core maintains a short term key to decrypt OR's information.

\textit{2) Network Slicing Establish.}
To initiate the establishment of the first hop in the circuit, $UE$ selects a link ID, denoted as $L_{1}$, and transmits a $create$ link request to $RAN1$. The payload of this request encapsulates the partial key $g^{x_{1}}$ derived from the Diffie-Hellman key exchange algorithm, and is encrypted specifically for $RAN1$. Upon receiving the request, $RAN1$ proceeds to decrypt the payload, validate its legitimacy, and respond with a $created$ link $L_{1}$ that comprises the complementary half-key $g^{y_{1}}$ associated with the Diffie-Hellman algorithm, as well as the hash of the negotiated key, $K_{1} = g^{x_{1} y_{1}}$. After $UE$ receives the response and validates the hash of the symmetric key, the link between $UE$ and $RAN1$ is established, with the session key between them being $K_{1}$.

Upon establishing the link, $UE$ transmits a $Relay$ link request $L_{1}$ to $RAN1$, which encapsulates the address of $RAN2$, the encrypted partial key $C_{RAN2} = E_{PK_{RAN2}}(g^{x_{2}})$ obtained from the Diffie-Hellman algorithm, and the encrypted target 5G core information $C_{Core} = E_{PK_{RAN2}}(T_{core})$. The entirety of the $Relay$ link request is secured using the session key $K_{1} = g^{x_{1} y_{1}}$ in conjunction with the 128-bit AES cipher operating in counter mode \cite{Tor}. $RAN1$ subsequently decrypts the payload, extracts $RAN2$'s address, and selects a new link ID, denoted as $L_{2}$. Additionally, $RAN1$ encrypts its router descriptor $D_{RAN1}$ using the 5G Core's public key $PK_{core}$, resulting in $E_{RAN1} = E_{PK_{core}}(D_{RAN1})$.

$RAN1$ then forwards a $create$ link request $L_{2}$ to $RAN2$, with the payload copied from the $Relay$ request and appended with $E_{RAN1}$. In response, $RAN2$ decrypts the payload, verifies its authenticity, and issues a $created$ link message $L_{2}$ containing the complementary half-key $g^{y_{2}}$ as well as the hash of the negotiated key $K_{2} = g^{x_{2} y_{2}}$. $RAN2$ also encrypts its router descriptor as $E_{RAN2} = E_{PK_{core}}(D_{RAN2})$ and encapsulates it alongside $E_{RAN1}$. Once $RAN1$ receives this response, it relays the information back to $UE$ by utilizing the link ID $L_{1}$.

\textit{3) 5G Core Connection.}
At the same time, $RAN2$ decrypts 5G core information, and initiates an NG setup request to establish a connection with the 5G core. The 5G core then commences the registration procedure for the 5G standalone (SA) architecture, ultimately establishing the 5G SA architecture. Once the uplink \& downlink traffic transfers are successfully completed, $UE$ can initiate the data transmission process.

\textit{4) Data Transmission.} For a two-hop circuit, the slice ID in the NSI ID is partitioned into three segments: $ID_{RAN1} || ID_{RAN2} || ID_{CORE}$. $UE$ encapsulates the information required by each RAN, denoted as $info_{i}$ = ($NSSAI$, $ID_{i}$, $Bearer\_ Context$, $Security \_ Info$, $RRC\_ Config$, $UE\_ Identity$, $Timestamp\_ SeqNum$, $\\Packet\_ Type$), and encrypts as: $E_{K1}(info_{RAN1}||$ $E_{K2}(info_{RAN2}|| ADDRESS_{CORE} ||ID_{CORE}||Data)).$
The last RAN, $RAN2$, must encapsulate the information required by the 5G core, defined as
$info_{core} =(Bearer Context$, $UE Identifier$, $Uplink\& Downlink \ Packet \ Counters$, $Timestamp$, $Security Parameters$, $Signal Quality Metrics$$)$.
The encrypted payload is transmitted, and each node decrypts the outermost layer, extracts the encrypted payload, and forwards it to the subsequent hop. This process continues until the final node decrypts the payload, retrieves the core network information, and transmits the data and $info_{core}$ to the core network for additional processing and forwarding.

Upon receiving the data, the 5G core network identifies the destination $UE$ or server based on the decoded data and $info_{core}$ by examining the $UE$'s session information or querying the target server for a specific service in the database. The core network selects an appropriate transmission path, considering factors such as the current network status, load balancing, and latency. Subsequently, the data is forwarded to the target UE or server, traversing multiple network nodes. Each node is responsible for executing respective processing tasks on the data, such as adjusting QoS and applying security policies. Finally, upon receiving the data, the target UE or server processes it according to the requested application or service. In this transmission process, the 5G core network ensures the efficient and secure transfer of data while adhering to specific service requirements and network policies.

\section{Security Analysis}

In this section, we explain how the proposed SPNS achieves secure network slice selection, anonymous authentication, and end-to-end authentication.

\textit{1) Secure Network Slice Selection:} In our design, we employ onion routing to ensure privacy protection in communication between users, RAN nodes, and the 5G core network. During transmission, user data is encrypted in multiple layers, with each layer corresponding to a RAN node. The encryption function of the $i$-th RAN node is represented by $E_{K_i}$, while its decryption function is represented by $D_{K_i}$. Each RAN node can only decrypt the encryption layer corresponding to it, ensuring end-to-end communication privacy.

To generate the symmetric key $K_i$ for each RAN node, we use the Diffie-Hellman (DH) key exchange protocol. Let $g$ be the generator and $p$ be a large prime number. The key negotiation process between the user and the RAN node is as follows: the user selects a random number $x$ and calculates $A=g^x \bmod{p}$; the RAN node selects a random number $y$ and calculates $B=g^y \bmod{p}$. The user and the RAN node then exchange $A$ and $B$, and calculate $K_U=B^x \bmod{p}$ and $K_R=A^y \bmod{p}$, respectively. Since $K_U=K_R=g^{xy} \bmod{p}$, a shared key $K$ is established between the user and the RAN node for encrypting and decrypting communication data.

Simultaneously, information related to the connection established by each node is encrypted using the node's public key, ensuring that only the relevant nodes can access and utilize the information for connection and data transition purposes. This provides an additional layer of security to the network, helping to ensure the confidentiality and integrity of the data being transmitted.

\textit{2) Anonymous Authentication:} In order to achieve anonymous authentication, the last RAN packages the services of both RANs to the 5G core, and the secondary RAN will not have direct contact or communicate with the 5G core. Therefore, the 5G core will not be able to communicate directly with the secondary RAN, and the secondary RAN will not be able to provide complete services to the 5g core alone. As long as no collusion between the 5G core and the secondary RAN, the 5G core cannot learn the identity of the mobile user.

\textit{3) End-to-End Encryption:} The network slices using encryption algorithms such as AES implements end-to-end encryption for data transmission. This ensures that only the recipient with the correct key can decrypt and obtain the original data during transmission. The encryption function can be represented as $E = E_{K_{AES}}(Data)$, and the decryption function as $D_{K_{AES}}(E)$, where $K_{AES}$ is the key for the AES encryption algorithm.

\section{Performance}
We conducted extensive experiments to measure the time cost associated with establishing network slice links under various conditions. The experiments were conducted on an OptiPlex 7490 AIO equipped with an Intel(R) Core(TM) i7-10700 CPU @ 2.90 GHz and 32 GB of RAM. We utilized the JSBN library for implementing RSA and ECC cryptographic operations. Each link cell had a size of 512 bytes, with a payload of 498 bytes for streams.

We conducted experiments to investigate the relationship between the size of the transmission data and connection time cost in SPNS. The transmission data sizes ranged from 1Mb to 14Mb, and each configuration was executed 100 times. We calculated the average running time across all iterations and analyzed the results.

\begin{figure}
    \centering
    \includegraphics[width=0.35\textwidth]{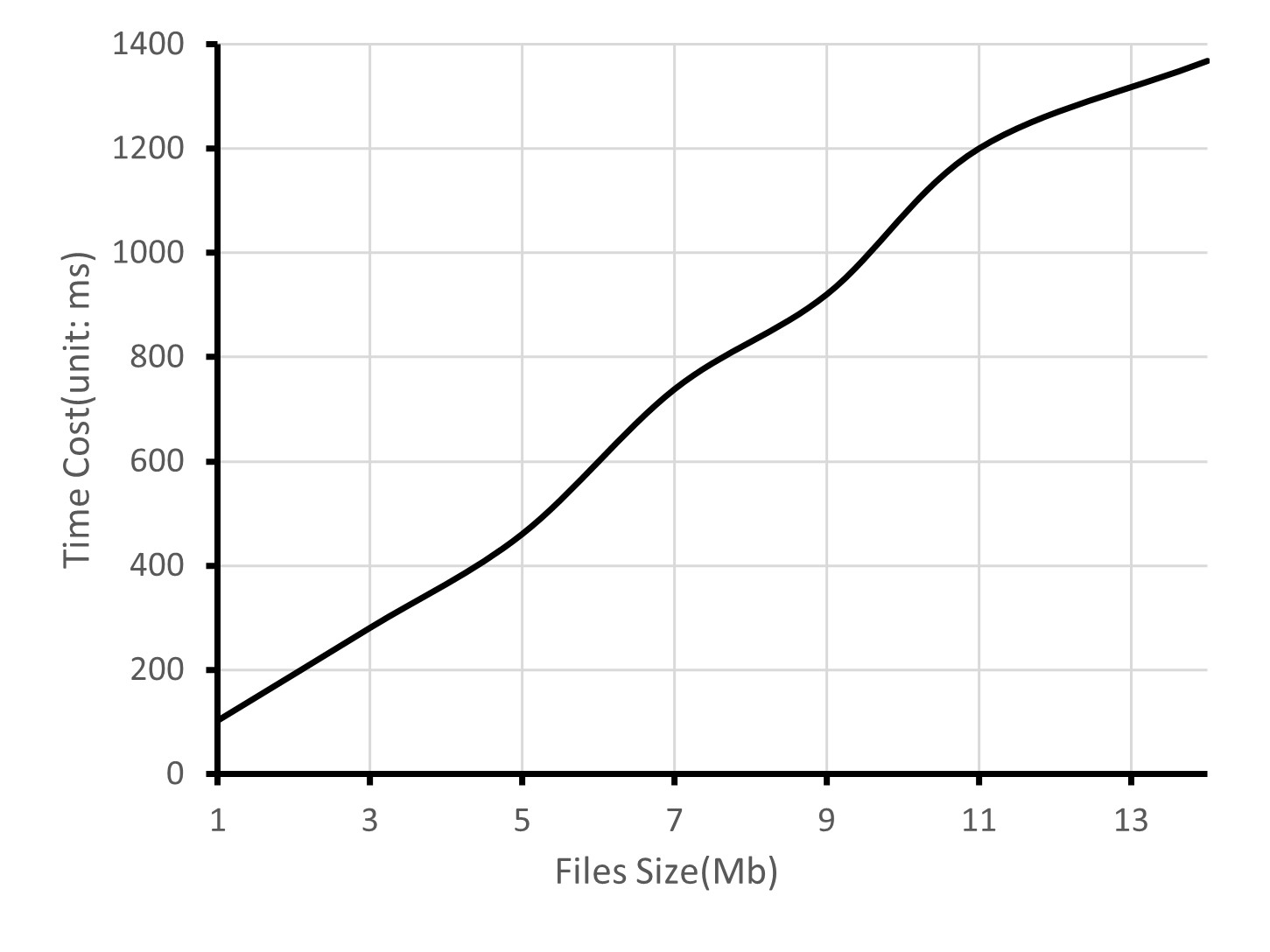}
    \caption{Time Cost of Increasing Nodes.}
    \label{cost}
\end{figure}

Our findings show a linear increase in connection time cost as the size of the transmission data in the network grows. We observed that the transmission speed was around 10 Mb/s, which is a bit slower than the unsecured transmission due to the added overhead of connection establishment, encryption, and decryption. Figure \ref{cost} illustrates the time cost of connections for different sizes of transmission data. These results suggest that SPNS is suitable for applications that require high-speed data transmission, while providing anonymity and privacy benefits for network communications.

\section{Conclusion}
In this paper, we have presented a novel protocol for the 3GPP 5G network slicing architecture that leverages the idea of onion routing to enhance privacy and security while maintaining efficiency in mobile communication. The proposed protocol conceals the identities of UE and RANs from the 5G core, preventing potential adversaries from tracking user data and compromising network operations. Furthermore, the design incorporates a multi-hop circuit and encryption techniques, enabling end-to-end secure communication between the UE, RANs, and the 5G core. By integrating onion routing and existing 5G components, the proposed protocol establishes a robust and privacy-preserving framework for data transmission in 5G networks, which can readily adapt to various network conditions, service requirements, and performance metrics. In the future work, we will explore potential optimizations and enhancements of the proposed protocol, addressing scalability, performance, and integration with emerging technologies and network paradigms.

\bibliographystyle{IEEEtran}
\bibliography{IEEEabrv,myrefs}

\end{document}